\documentclass[review]{elsarticle}
\pdfoutput=1
\usepackage[utf8]{inputenc}
\usepackage{graphicx}
\usepackage{amsmath,amstext,amssymb,color,amsthm,float}
\usepackage{subfig}

\begin{document}

\begin{frontmatter}

 \author[lab3]{Elena~Loli~Piccolomini}
 \author[lab1]{Fabiana~Zama}
 \address[lab1]{Department of Mathematics, University of Bologna}
 \address[lab3]{Department of Computer Science and Engineering, University of Bologna}

%\date{March 2020}
\title{Preliminary analysis of  COVID-19 spread in Italy with an adaptive SEIRD model }

\begin{abstract}

In this paper we propose a Susceptible-Infected-Exposed-Recovered-Dead (SEIRD) differential model for the analysis and forecast of the COVID-19 spread in some regions of Italy,  using the data from the Italian Protezione Civile from February 24th 2020. In this  study investigate an adaptation of the model.  
Since several restricting measures have been imposed by the Italian government at different times, starting from  March 8th 2020, we 
propose a modification of  SEIRD by introducing a  time dependent transmitting rate. In the numerical results we report  the maximum infection spread for the three Italian regions firstly affected by the COVID-19 outbreak(Lombardia, Veneto and Emilia Romagna). This approach will  be successively extended to other Italian regions, as soon as more data will be  available. 
\end{abstract}
\end{frontmatter}

%\maketitle

\section{Introduction}
The recent diffusion of the COVID-19 Corona virus has renewed the interest of the scientific and political community in the  mathematical models for epidemic. Many researchers are making efforts for proposing new refined models to analyse the present situation and predict possible future scenarios.\\
With this paper we hope to contribute to the ongoing research on this topic and to give a practical instrument for a deeper comprehension of the virus spreading features and behaviour.\\
We consider here deterministic models based on a system of initial values problems of Ordinary Differential Equations (ODEs).
This theory has been studied since about one century by W.O. Kermack and A. G. MacKendrick \cite{KMK27} that proposed the basic Susceptible-Infected-Removed (SIR) model. The SIR model and its later modifications, such as Susceptible-Exposed-Infected-Removed (SEIR) \cite{Ebola13} are commonly used by the epidemic medical community in the study of outbreaks diffusion.% For example, in the case of Ebola 
In these models, the population is divided into groups. For example,  the SIR model  groups are:  Susceptible who can catch the disease, Infected who have the disease and can spread it, and Removed those who have either had the disease, or are recovered, immune or isolated until recovery.  The SEIR model proposed by Chowell et al. \cite{Ebola2004} considers also  the Exposed group: 
containing  individuals who are in the incubation period. \\
%Each group is represented by a continuous function. 
The evolution of the Infected group depends on a key parameter, usually denoted as R0, representing the basic reproductive rate. The value of R0 can be inferred, for example, by epidemic studies or by statistical data from literature or it can be calibrated from the available data. In this paper we use the available data for determining the value of R0 best fitting the data.\\
Compared to previous outbreaks, such as SARS-CoV or MERS-CoV \cite{lancet2}, when the disease had been stopped after a relatively small number of infected people,  we are now experimenting a different situation. Indeed  the number of infected people grows exponentially, and
apparently, it can be  stopped only by  a complete lockdown of the affected areas, as  evidenced by the  COVID-19 outbreak in the Chinese city of Wuhan in December 2019. \\
Analogously, in the Italian case, in order to limit the virus diffusion all over the Italian area,  the government has started to impose
more and more  severe restrictions  since March 6th 2020. 
Hopefully, these measures  will affect  the spread of the COVID-19 virus reducing the number of infected people and 
the value of the parameter R0. \\
The introduction of different levels of lockdown  require an
adaptation of the standard epidemic models to this new situation.
Some examples about the   Chinese outbreak can be found in  
\cite{lancet1,lancet2,chinese1}. Concerning the 
Italian situation, which is currently evolving, 
it is possible to model the introduction of restricting measures
by introducing a non constant infection rate \cite{Fanelli20}.
In this paper we propose to represent the  infection rate as a 
piecewise function,  which reflects the changes of external conditions.  The parameter R0, which is proportional to the infection rate, becomes a time dependent parameter $R_t$ 
which follows a different trend  each time the external conditions change,
depending on the particular situation occurring in that period.
 For example, if new restrictions are applied to the population movements at time $t_1$, we can hopefully argue that $R_t$ starts to decrease
 when $t>t_1$.\\
 Finally, we believe that relevant information is not only Infected but also Recovered and Dead numbers we  modified SEIR model by splitting the Removed population into Recovered and Dead.\\
In section \ref{model} we describe the details of  SEIRD model with constant and with time dependent infection rate SEIRD(rm).
Finally in section \ref{Nres}
we test the model on some regional aggregated data published by the Protezione Civile Italiana \cite{SitoPC}.
\section{Numerical model and methods \label{model}}
The SIR model proposed by Kermack and A. G. MacKendrick \cite{KMK27} divides the population in three groups: Susceptible (S), Infected (I) and Recovered(R). The equations relating the groups are the followings:
\begin{eqnarray*}
    \frac{dS}{dt}&=&\frac{\beta}{N} SI\\
    \frac{dI}{dt}&=&\frac{\beta}{N}SI - \gamma I\\
     \frac{dR}{dt}&=& \gamma I
\end{eqnarray*}
where $N$ is the total population, $\beta$ is the infection rate, a coefficient accounting for the susceptible people get infected by infectious people and $\gamma$ is the parameter of infectious people which become resistant  per unit time. A more refined model is the SEIR model where a new compartment E representing the exposed individuals that are in the incubation period is added.

The resulting equations in the SEIR model are the following:
\begin{eqnarray*}
    \frac{dS}{dt}&=&\frac{\beta}{N} SI\\
    \frac{dE}{dt}&=&\frac{\beta}{N} SI -\alpha E\\
    \frac{dI}{dt}&=& \alpha E - \gamma I\\
     \frac{dR}{dt}&=& \gamma I
\end{eqnarray*}
where $\alpha$ represents the incubation rate.
The difference between the exposed (E) and infected (I) is that the former have contracted the disease but are not infectious, and the latter can spread the disease.  SEIR  has been  used to model breakouts, such as Ebola in Congo and Uganda \cite{Ebola2004}. In \cite{Ebola13} the equations are modified by adding the quarantine and vaccination coefficients. In our case, unfortunately, vaccination is not available.
A further model, the SIRD, considering the group of  Dead (D) in place of the Exposed is analysed in \cite{Fanelli20} for the forecast of COVID 19 spreading.

In this paper we propose a SEIRD model accounting for five different groups, Susceptible, Exposed, Infected, Recovery and Dead. The system of equations is given by:

\begin{align}
&    \frac{dS}{dt}=\frac{\beta}{N} SI\nonumber \\
&   \frac{dE}{dt}=\frac{\beta}{N} SI -\alpha E \nonumber \\
\label{eq:seird} &   \frac{dI}{dt}= \alpha E - (\gamma_R+\gamma_D) I  \\
&     \frac{dR}{dt}= \gamma_R I \nonumber\\
&     \frac{dD}{dt}= \gamma_D I \nonumber 
\end{align}
In order to consider the restrictions imposed by the Italian government since March 8th 2020, we have partitioned the whole time interval $[0,T]$, considered for the integration of \eqref{eq:seird},  into two sub-intervals: $[0,t_0]$ and $(t_0,T]$ where $t_0$ 
corresponds to time when  the restrictions start to  produce a valuable change in the data trend. 
Moreover, since the applied restrictions should decrease the number of contacts between Infected and Susceptible, we model the coefficient $\beta$ in \eqref{eq:seird} as a decreasing time dependent function $\beta(t)$. 
%If we think to $X$ as the  random variable counting the meets between Infected and Susceptible and $beta_t$
 A similar model for the  infection rate of  SEIR equations can be found  in \cite{Fanelli20}, where the function is assumed to have a decreasing exponential form. 
 However, observing the data trend, we believe that $\beta_t$ has a smoother decreasing behavior 
 and we choose to model it as a decreasing rational function: 
\begin{equation}
    \beta(t)=\left \{ \begin{array} {ll}
    \beta_0 & if \ t < t_0 \\
    \beta_0 \left (1-\rho (t-t_0)/t) \right )&  otherwise
    \end{array}
    \right., \ \ \ \rho \in (0,1). 
    \label{eq:beta_t}
\end{equation}
In the present work we use a constant value $\rho=0.75$ but we might calibrate it in future.
By substituting $\beta(t)$ \eqref{eq:beta_t} in  the $S$ and $E$ equations in \eqref{eq:seird} we obtain SEIRD rational model
SEIRD(rm). \\
We calibrate the parameters  of SEIRD and SEIRD(rm) 
by solving  non-linear least squares problems with positive constraints.
For example, in the SEIRD model \eqref{eq:seird},  we define the 
function  $\mathbf{u}(t)=(S(t),E(t),I(t),R(t),D(t))$, depending on the  vector  of parameters $\mathbf{q}=(\beta, \alpha, \gamma_R, \gamma_D)$, 
and the vector $\mathbf{y}$ of the acquired data at given times $t_i, i=1, \ldots n$. Let $F(\mathbf{u,q})$ be  the function computing the numerical solution $\mathbf{u}$ of the differential system \eqref{eq:seird},  the estimation
of the parameter $\mathbf{q}$  is obtained solving the following non linear least squares problem:
 \begin{eqnarray}
 \min_{\mathbf{q}} \frac{1}{2}\| F(\mathbf{u,q}) - \mathbf{y} \|_2^2 
 \label{eq:minq_c}
 \\
 \mathbf{q} \geq 0 \nonumber
 \end{eqnarray}
 where we  introduce  positivity constraints on $\mathbf{q}$.
 The constrained optimization problem is solved with a trust-region based method implemented in the \texttt{lsqnonlin} Matlab function. For further details about the optimization problem for identification parameters in differential problems see for example \cite{IFIP17}.
%----------------------------------------------------------------------
\section{Numerical Results \label{Nres}}
In this section we report the results obtained by using the SEIRD model 
to monitor the Covid-19  outbreak in Italy during the period 24/02/2020-20/03/2020.
The epidemic spread started on February 21st affecting  the northern regions. Lombardia in particular 
registered the first epidemic outbreak followed by the Veneto region,  Emilia Romagna  
the other Italian regions. \\
Since initially each region applied different containment measures to
some restricted areas  at different times,  we chose to calibrate the SEIRD model on each region separately.
This study considers Lombardia, Veneto and Emilia Romagna regions for which the largest amount of
meaningful data have been collected in the GitHub repository \cite{SitoPC}.  \\
The purpose is to develop a model calibration and simulation method to be eventually extended 
 to all the other Italian regions reached by the epidemic spread.
%The simulations are run in different phases that are explained in the subsequent paragraphs.\\
All computations are performed using Matlab R2019b 2,9 GHz Intel Core i7 quad-core 16 GB ram. 
The method consist of two main steps: 
\begin{itemize}
\item {\bf Identification}. In this step the model parameters are estimated  by means of the  {\tt trust-region-reflective} algorithm implemented in the {\tt lsqnonlin} matlab function.
\item {\bf Simulation}. This step applies the SEIRD-SEIRD(rm) models with the identified parameters in order to monitor the process for a longer time (up to  240  days). The differential system is solved applying the  {\tt ode45} matlab function with the following initial condition:
$S(0)=N$,$E(0)=I(0)=I_{init}$,$R(0)=D(0)=0$ where the value $I_{init}$ corresponds to the Infected individuals in the first measurement day.
\end{itemize}
\subsection{SEIRD simulation}
\begin{figure}[!hbtp]
\centering
\includegraphics[width=0.45\textwidth]{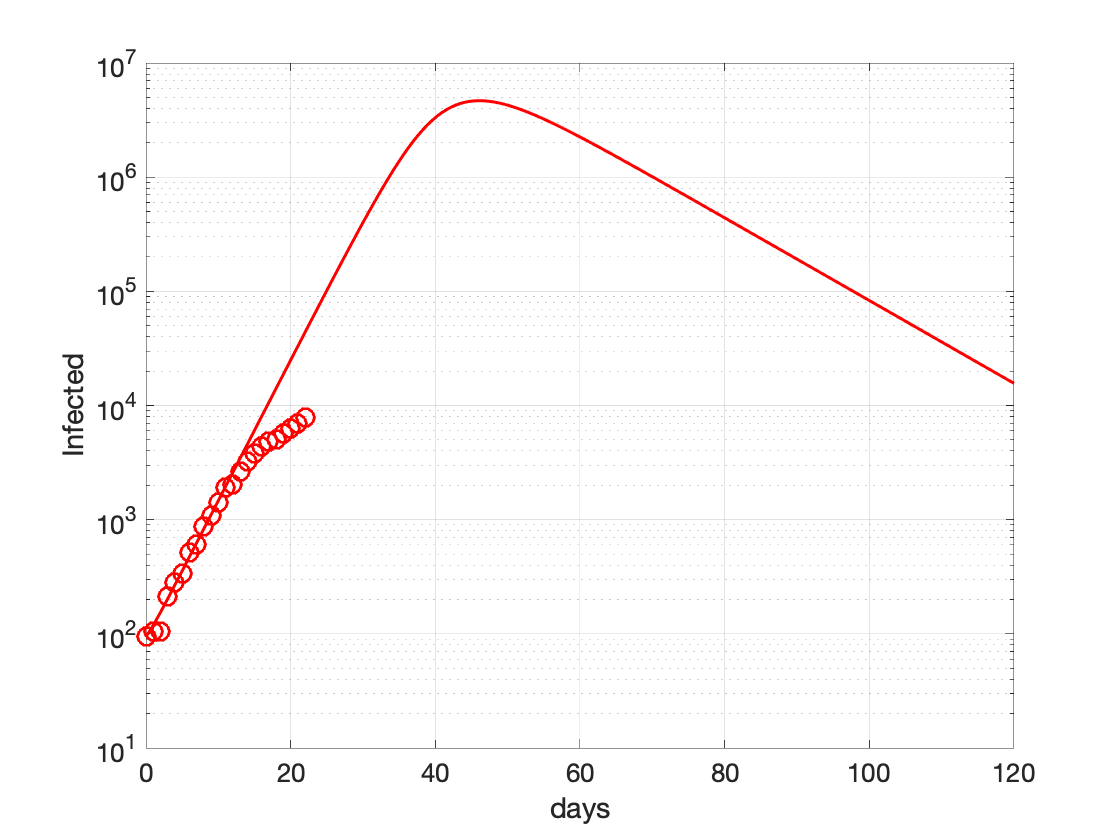}
\includegraphics[width=0.45\textwidth]{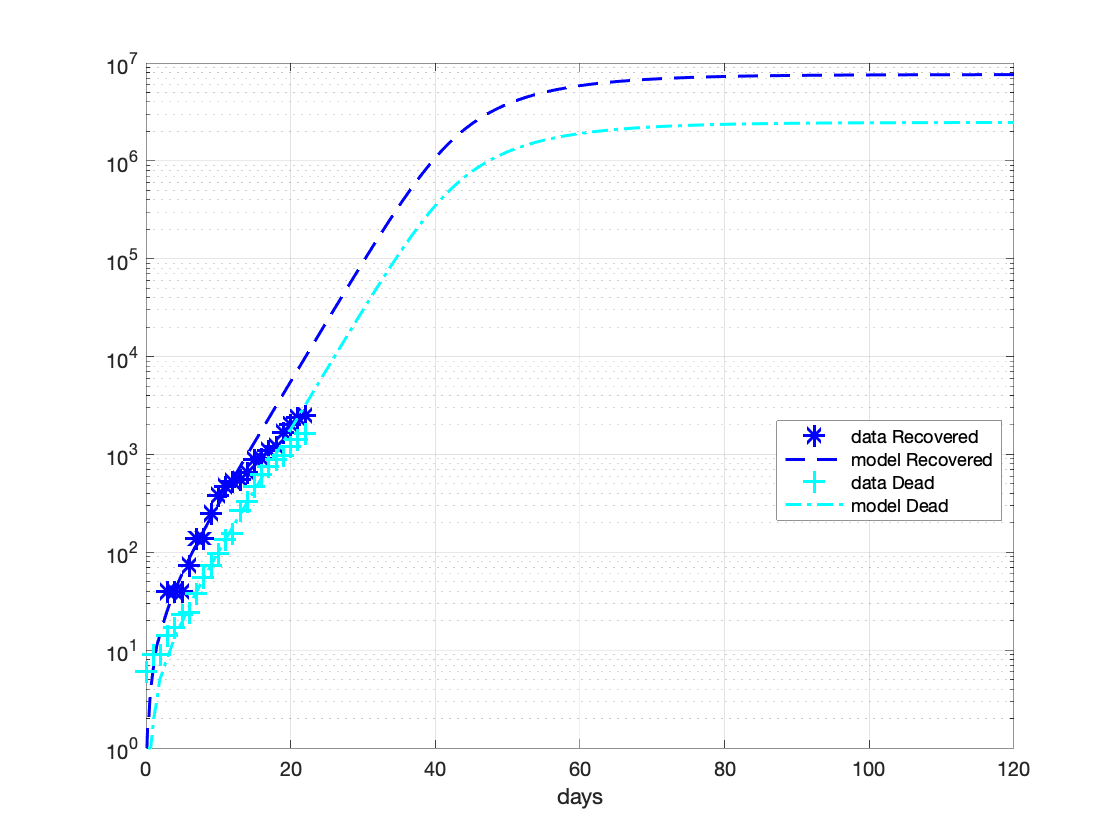}
\caption{Infect, Recovered Dead model based on  Lombardy data S1.}
\label{fig:Lo10}
\end{figure}

\begin{figure}[!hbtp]
\centering
\includegraphics[width=0.45\textwidth]{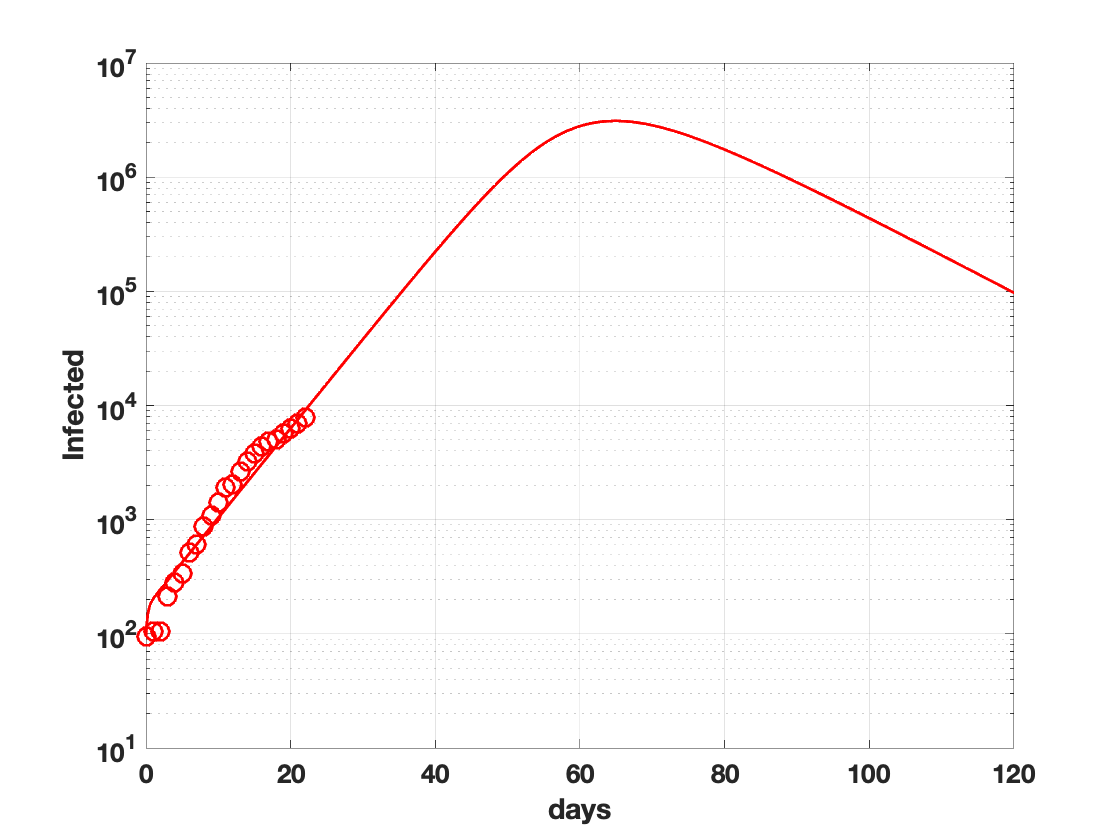}
\includegraphics[width=0.45\textwidth]{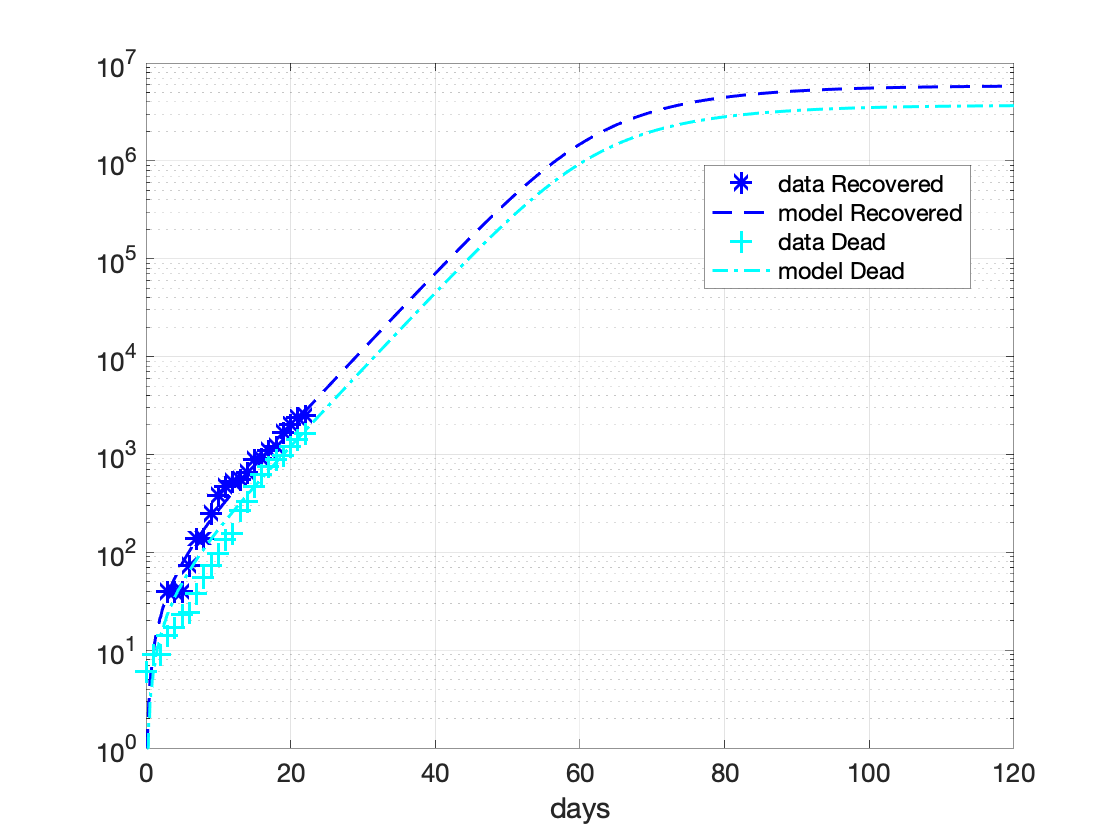}
\caption{Infect, Recovered Dead model based on  Lombardy data S3.}
\label{fig:Lo23}
\end{figure}

\begin{figure}[!hbtp]
\centering
\includegraphics[width=0.45\textwidth]{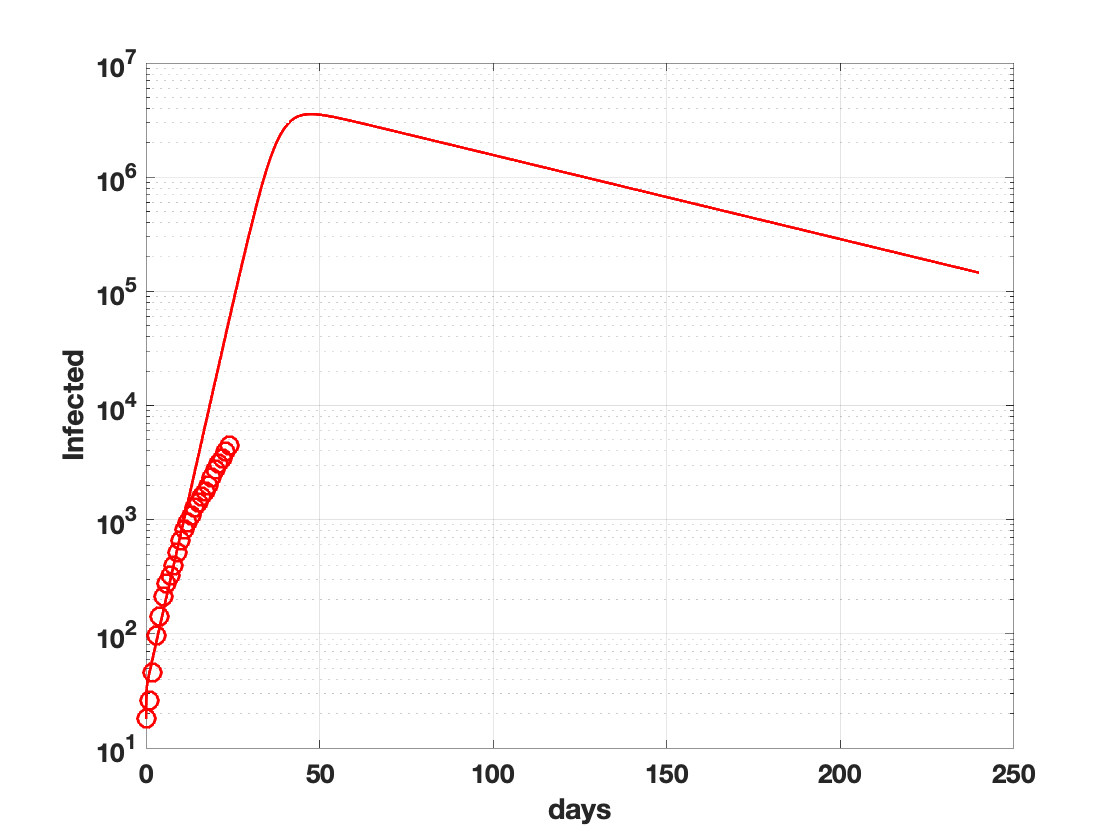}
\includegraphics[width=0.45\textwidth]{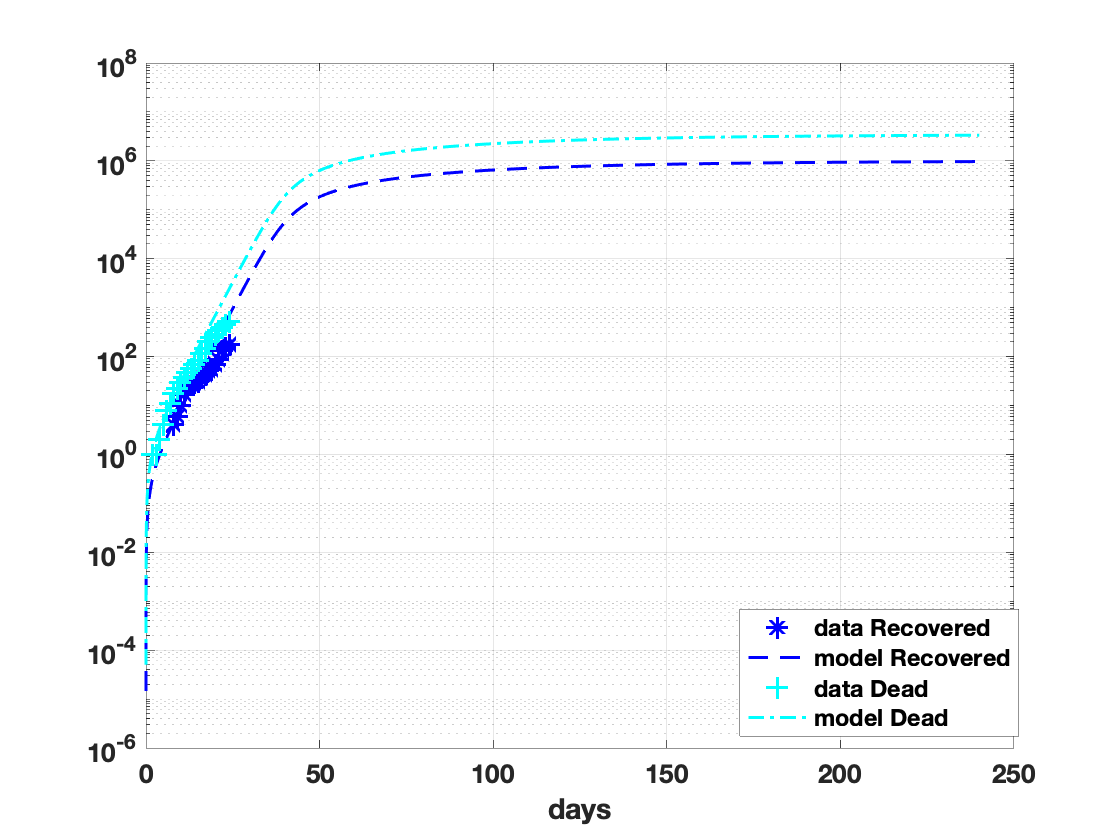}
\caption{Infect, Recovered Dead model based on  Emila Romagna data S1.}
\label{fig:ER10}
\end{figure}

\begin{figure}[!hbtp]
\centering
\includegraphics[width=0.45\textwidth]{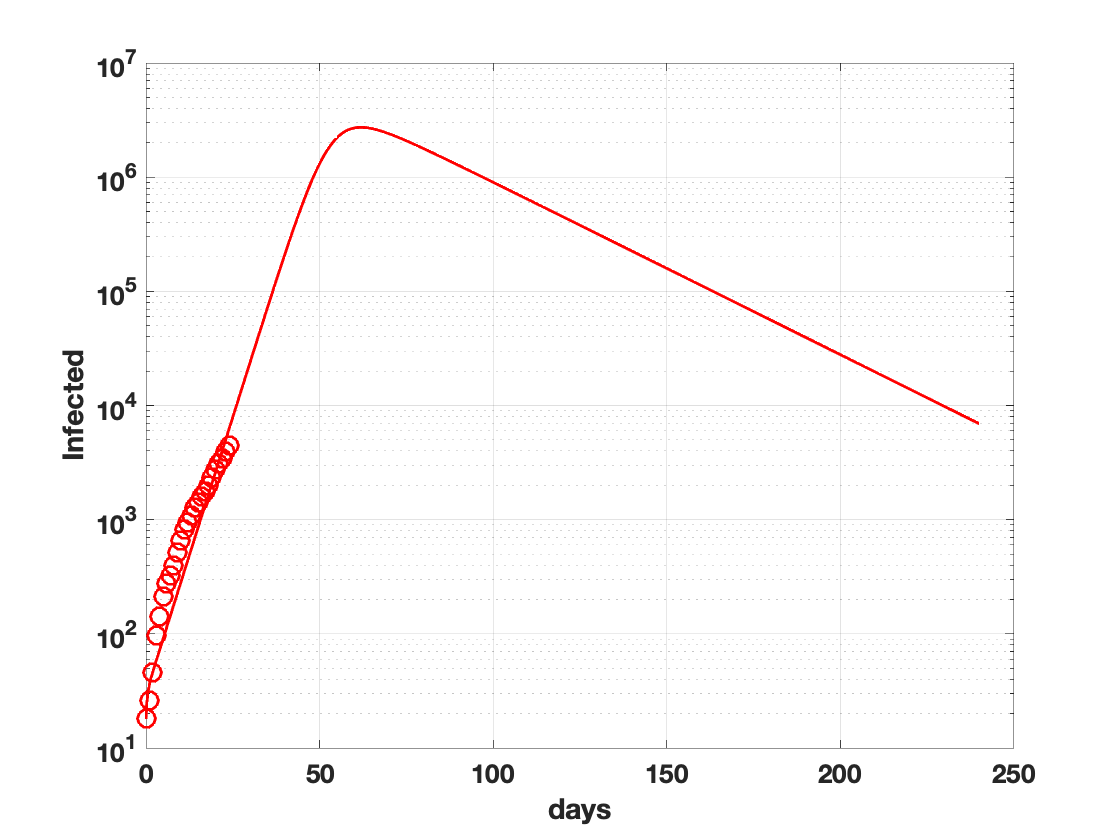}
\includegraphics[width=0.45\textwidth]{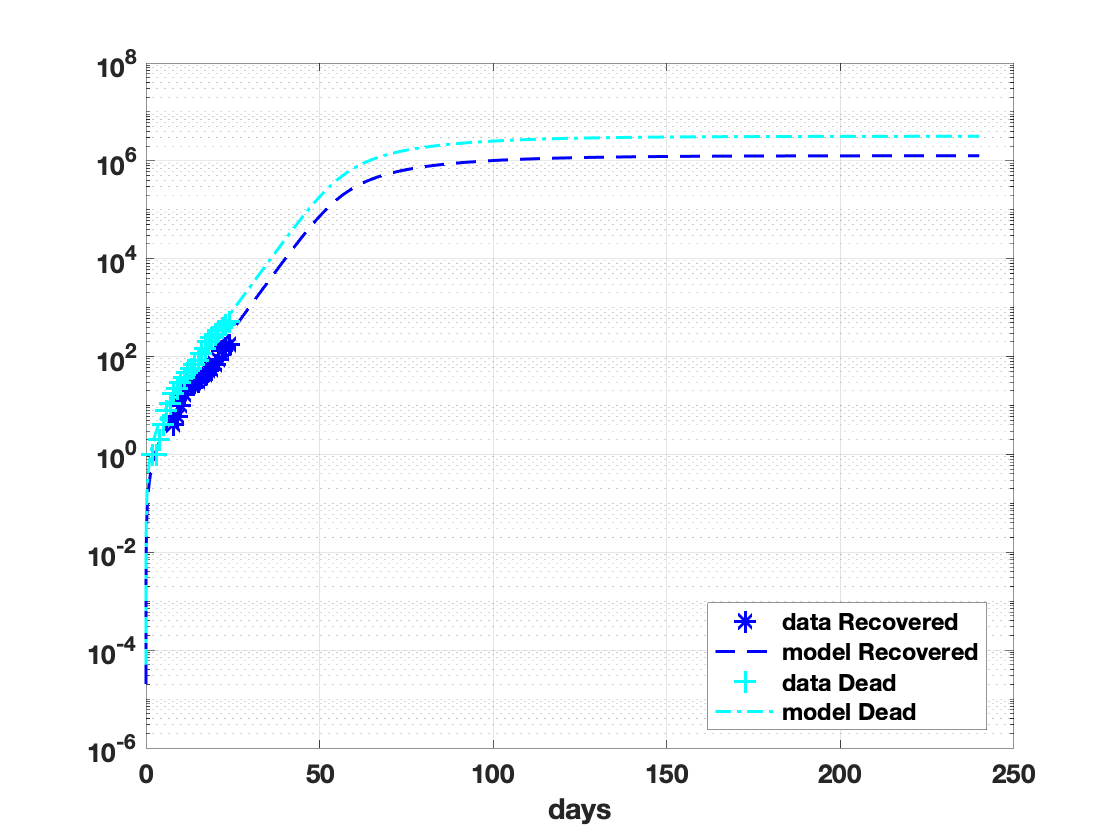}
\caption{Infect, Recovered Dead model based on  Emila Romagna data S3.}
\label{fig:ER23}
\end{figure}

\begin{figure}[!hbtp]
\centering
\includegraphics[width=0.45\textwidth]{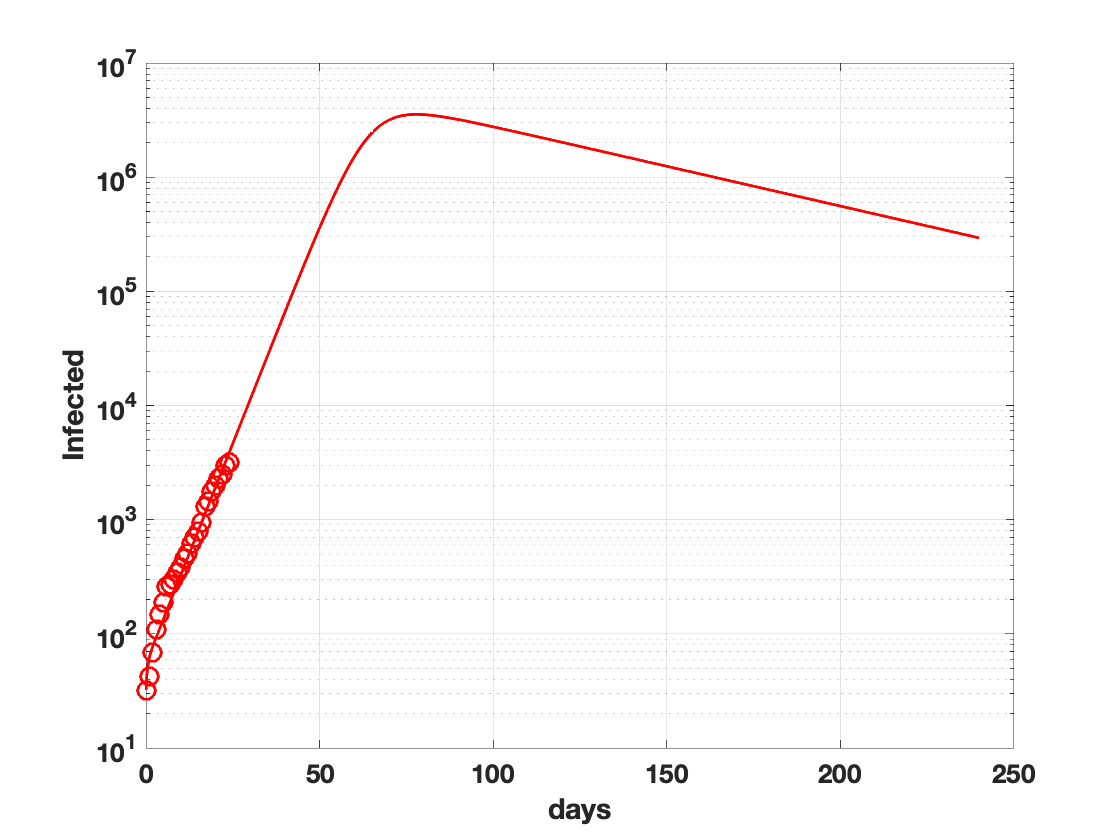}
\includegraphics[width=0.45\textwidth]{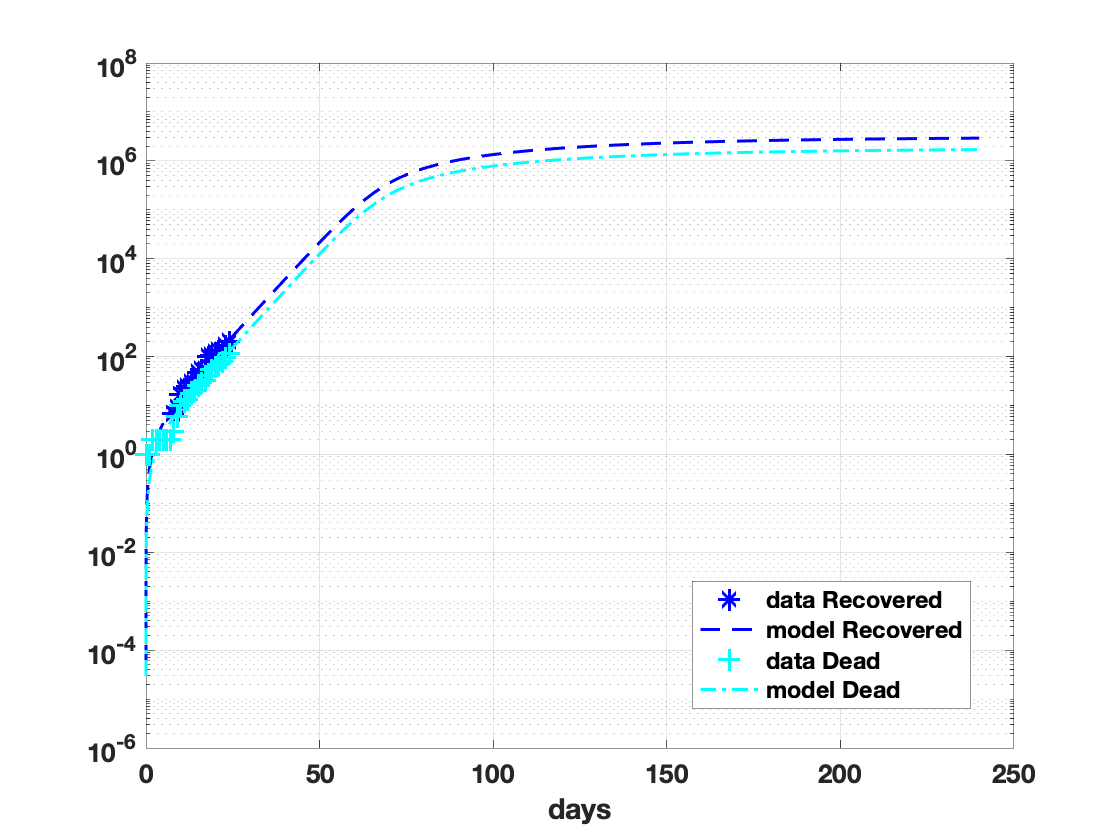}
\caption{Infect, Recovered Dead model based on  Veneto data S3.}
\label{fig:Ve23}
\end{figure}

In this section we consider the  SEIRD model \eqref{eq:seird} and use the following different measurement subsets from \cite{SitoPC} relative to Lombardia region:
\begin{itemize}
\item {\bf S1} 10 days measurements: 24/02/2020-04/03/2020 
\item {\bf S2} 18 days measurements: 24/02/2020-12/03/2020
\item {\bf S3} 23 days measurements: 24/02/2020-17/03/2020
\end{itemize}
for identification of the parameters and than apply the identified parameters to model the Infected-Recovered-Dead populations up to June 22nd (120 days). Besides population plots in figures \ref{fig:Lo10} ,\ref{fig:Lo23} we collect some meaningful quantitative information about the model parameters (table \ref{tab:LoPars}) and the peak values for Infected, Recovered and Dead populations (table \ref{tab:LoNums}).\\
%-----------------------------------------------------------------------------     
\begin{table}[!h]
\begin{center}
\begin{tabular}{c|ccccc}
Measured days & $\beta$ & $\alpha $ & $\gamma_R $ & $\gamma_D$ & $R_0$ \\
\hline
  S1 & 0.74 &  0.28 &  0.0632 & 0.0205 & 8.8 \\
  S2 & 0.29 & 3.28 &  0.0473 & 0.0272 & 3.9 \\
  S3 & 0.29 & 3.22 & 0.0560 &  0.0354 & 3.1 \\
  \hline
\end{tabular}
\caption{SEIRD Model parameters, Lombardia Region}
\label{tab:LoPars}
\end{center}
 \end{table}
 %-----------------------------------------------------------------------------
\begin{table}[!h]
\begin{center}
\begin{tabular}{c|cc|cc|cc}
\hline
Data &   \multicolumn{2}{c|}{Peak Infected} & \multicolumn{2}{c|}{Peak Recovered}  & \multicolumn{2}{c}{Peak Dead} \\
   data    &            day & Number (\%) &  day &Number (\%) &  day& Number (\%) \\
 \hline
 S1   &  46 & 4674665 (46.4\%) & 120 &7601989 (75.4\%) & 120 & 2466103 (24.5 \%)  \\
 S2   &  60 & 3931157 (39.0\%) &120 & 6190842 (61.4\%) & 120 &3554881 (35.2\%) \\
 S3   &  65 & 3108534 (30.8\%) & 120 & 5792528 (57.4\%) & 120 & 3664115 (36.3\%) \\
\hline
 %S2
 \hline 
 \end{tabular}
 \caption{SEIRD, Infected, Recovered Dead values. Lombardia region.}
 \label{tab:LoNums}
 \end{center}
 \end{table}
 %-----------------------------------------------------------
The results can be summarized as follows:
\begin{itemize}
\item The maximum Infected population is reached on 
April 9th 2020 for S1, April 23rd 2020 for S2 and April 28th 2020 for S3.
\item Concerning the values of the model parameters we have that
the transmitting rate $\beta$ can be estimated as $\beta_e=0.3$, 
the incubation rate $\alpha$ is approximately $\alpha_e=3$ while the 
recovery rate is approximately $(\gamma_R)_e = 0.06$
and the dead rate can be approximated as is approximately $(\gamma_D)_e = 0.04$. 
\end{itemize}
The following conclusions can be drawn:
\begin{itemize}
\item The SEIRD parameters, computed from the first 10 days measurements, do not model properly the measurements. Indeed the plots reported in figure \ref{fig:Lo10}  show that data of IRD populations have a slower increase rate, compared to the model predictions. 
\item  Considering  the parameters obtained from S3 we observe that 
even if the data fit is improved, the model does not  seem to reproduce 
properly the increasing slope of the Infected data, hence the previsions from this model may not be entirely reliable. 
\end{itemize}
Concerning Emilia Romagna region we have similar results. The plots in figures \ref{fig:ER10},\ref{fig:ER23} show that the Infected population increase rate does not reproduce the data correctly hence the peak values reported in table \ref{tab:ERNums}
may not be accurate. \\
  %ER
\begin{table}[!htp]
\begin{center}
\begin{tabular}{c|cc|cc|cc}
\hline
Data &   \multicolumn{2}{c|}{Peak Infected} & \multicolumn{2}{c|}{Peak Recovered}  & \multicolumn{2}{c}{Peak Dead} \\
   data    &            day & Number (\%) &  day &Number (\%) &  day& Number (\%) \\
 \hline
S1 & 48 & 3565588 ( 80.0) & 240 & 968332 ( 21.7) & 240 & 3345752 ( 75.0) \\
S2& 57 & 3207633 ( 71.9) & 240 & 1217341 ( 27.3) & 240 & 3191080 ( 71.6) \\
S3 &  62 & 2729893 ( 61.2) & 240 & 1272248 ( 28.5) & 240 & 3179219 ( 71.3) \\
\hline
 %S2
 \hline 
 \end{tabular}
 \caption{SEIRD, Infected, Recovered Dead values, Emilia Romagna region. }
 \label{tab:ERNums}
 \end{center}
 \end{table}
%-------------------------------------------------------------
The SEIRD model seems to be more accurate for  Veneto region where the plots in figure \ref{fig:Ve23} reproduce the data quite accurately. The peak values reported in table
\ref{tab:VeNums} represent the largest percentage of Infected and the longest time required to reach the peak.
 %Veneto
\begin{table}[!htp]
\begin{center}
\begin{tabular}{c|cc|cc|cc}
\hline
Data &   \multicolumn{2}{c|}{Peak Infected} & \multicolumn{2}{c|}{Peak Recovered}  & \multicolumn{2}{c}{Peak Dead} \\
   subset   &            day & Number (\%) &  day &Number (\%) &  day& Number (\%) \\
 \hline
S1 & 71 &4024199 ( 82.0) & 240 & 2107711 ( 43.0) & 240 & 1966666 ( 40.1)  \\
S2&  77 &3661977 ( 74.6) & 240 & 2835238 ( 57.8) & 240 & 1685415 ( 34.4) \\
S3 & 78 & 3547164 ( 72.3) & 240 & 2913752 ( 59.4) & 240 & 1698521 ( 34.6) \\
\hline
 %S2
 \hline 
 \end{tabular}
 \caption{SEIRD, Infected, Recovered Dead values, Veneto region. }
 \label{tab:VeNums}
 \end{center}
 \end{table}
%==================================================
%----------------------------------------------------------------------------
%--------------------------------------------------------------
\subsection{SEIRD(rm) calibration and simulation}
\begin{figure}[!hbtp]
\centering
\includegraphics[width=0.3\textwidth]{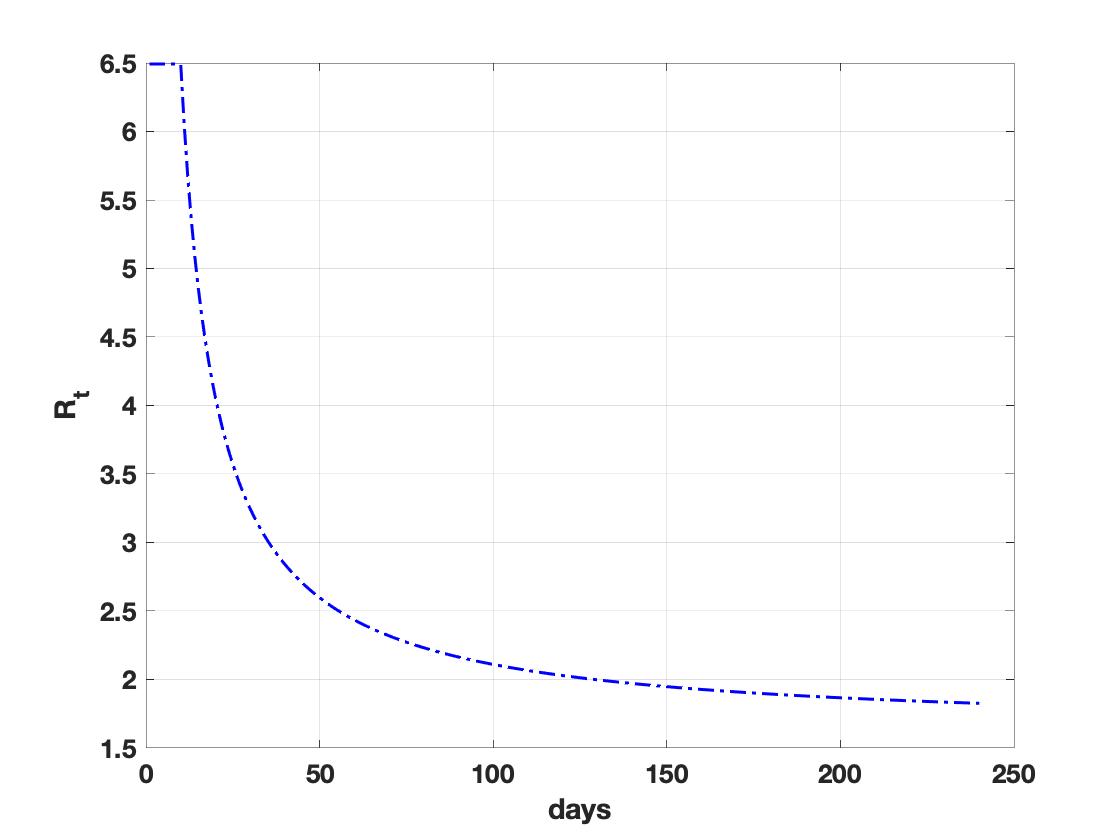}
\includegraphics[width=0.3\textwidth]{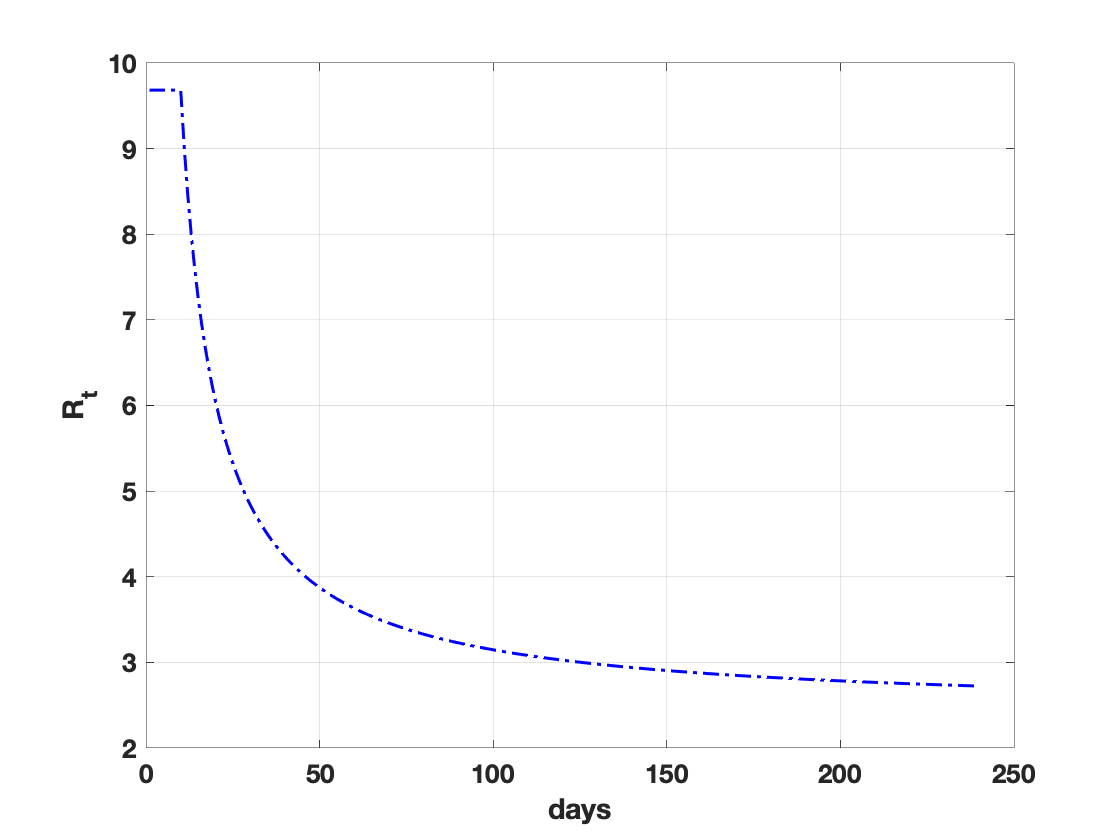} 
\includegraphics[width=0.3\textwidth]{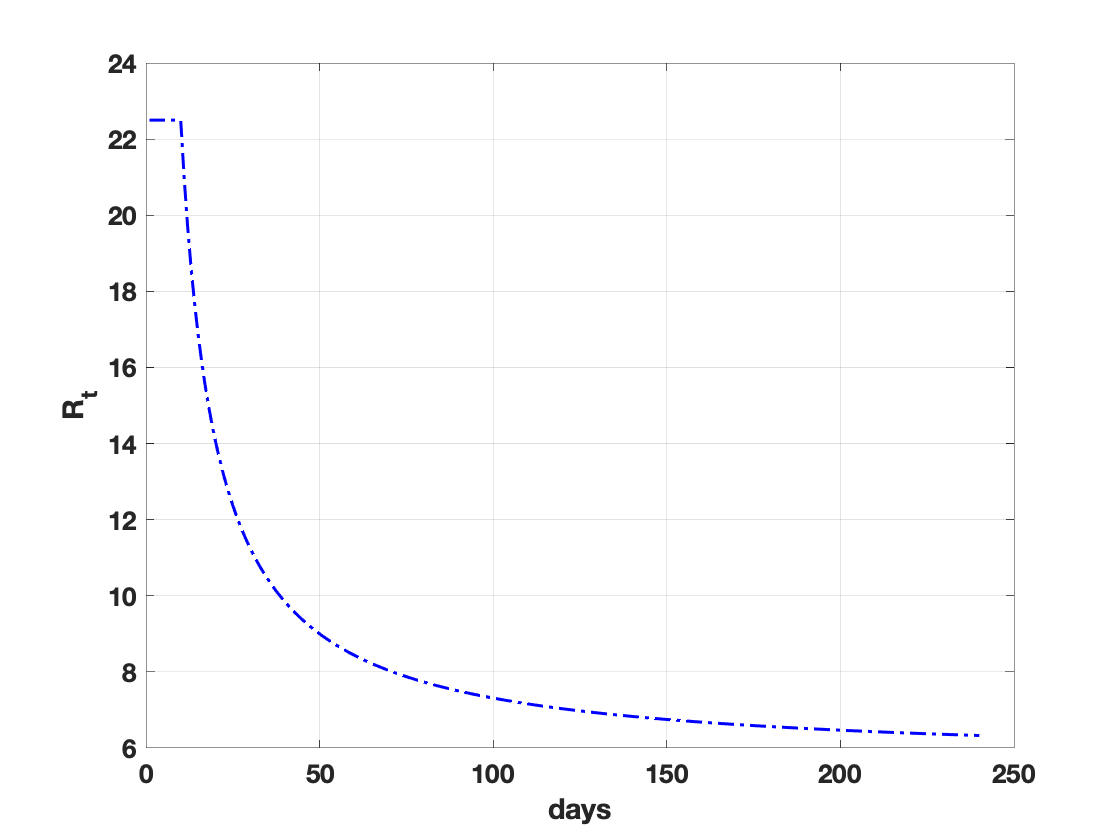} \\
Lombardia \hspace{1cm} Emilia Romagna \hspace{1cm} Veneto
\caption{Reproduction Rate $R_t$ of  SEIRD(rm) model.}
\label{fig:Rt}
\end{figure}

\begin{figure}[hbtp]
\centering
\includegraphics[width=0.45\textwidth]{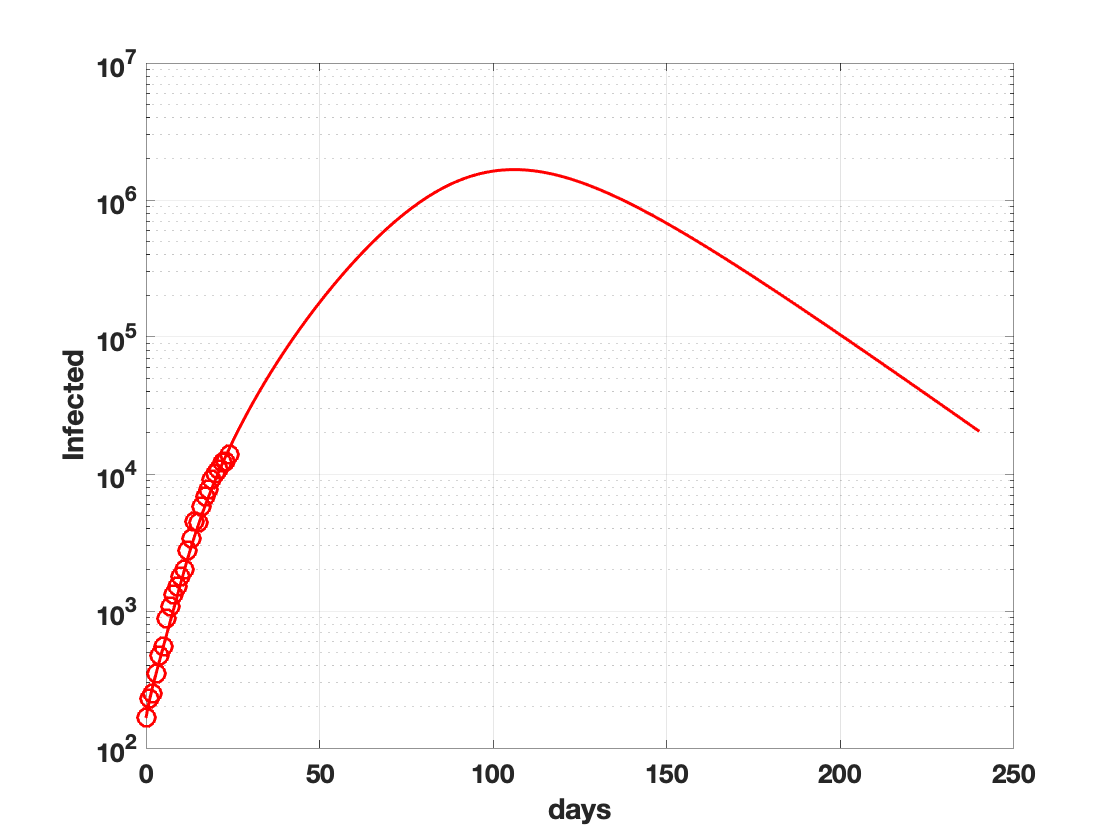}
\includegraphics[width=0.45\textwidth]{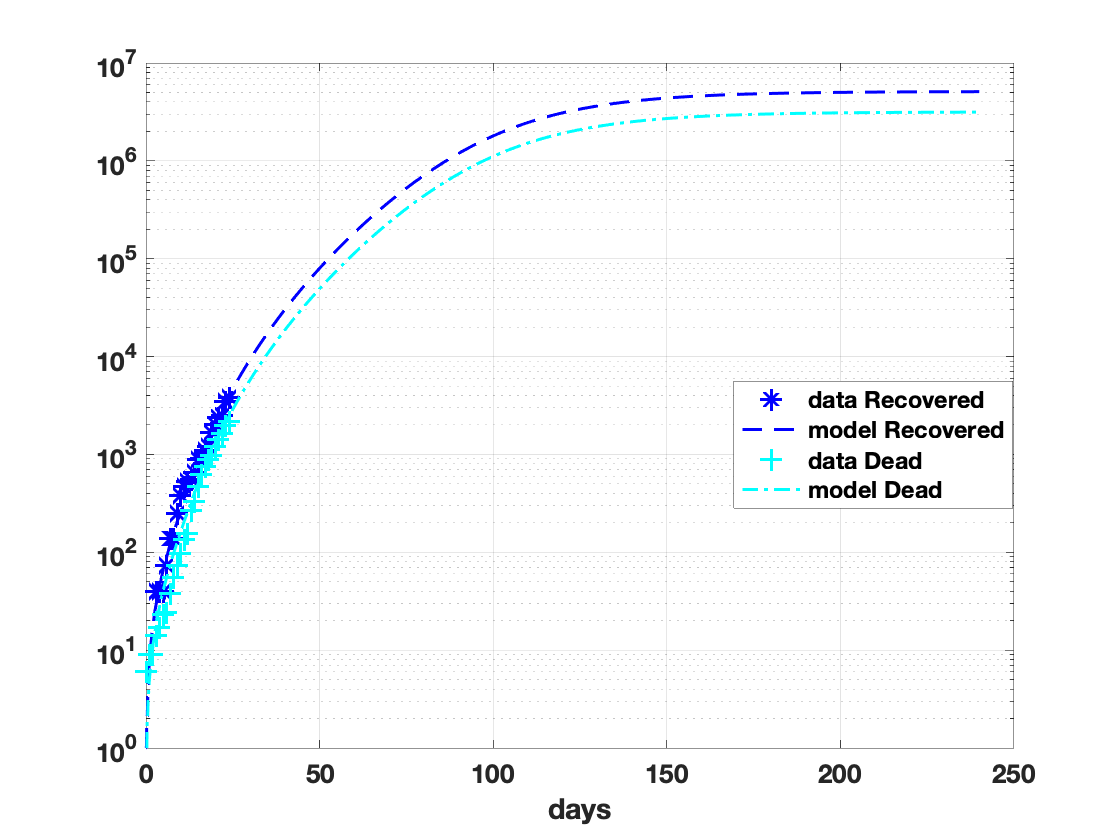}
\caption{Infect, Recovered Dead model based on  Lombardia region data (SEIRD(rm)).}
\label{fig:LoMod1}
\end{figure}

\begin{figure}[hbtp]
\centering
\includegraphics[width=0.45\textwidth]{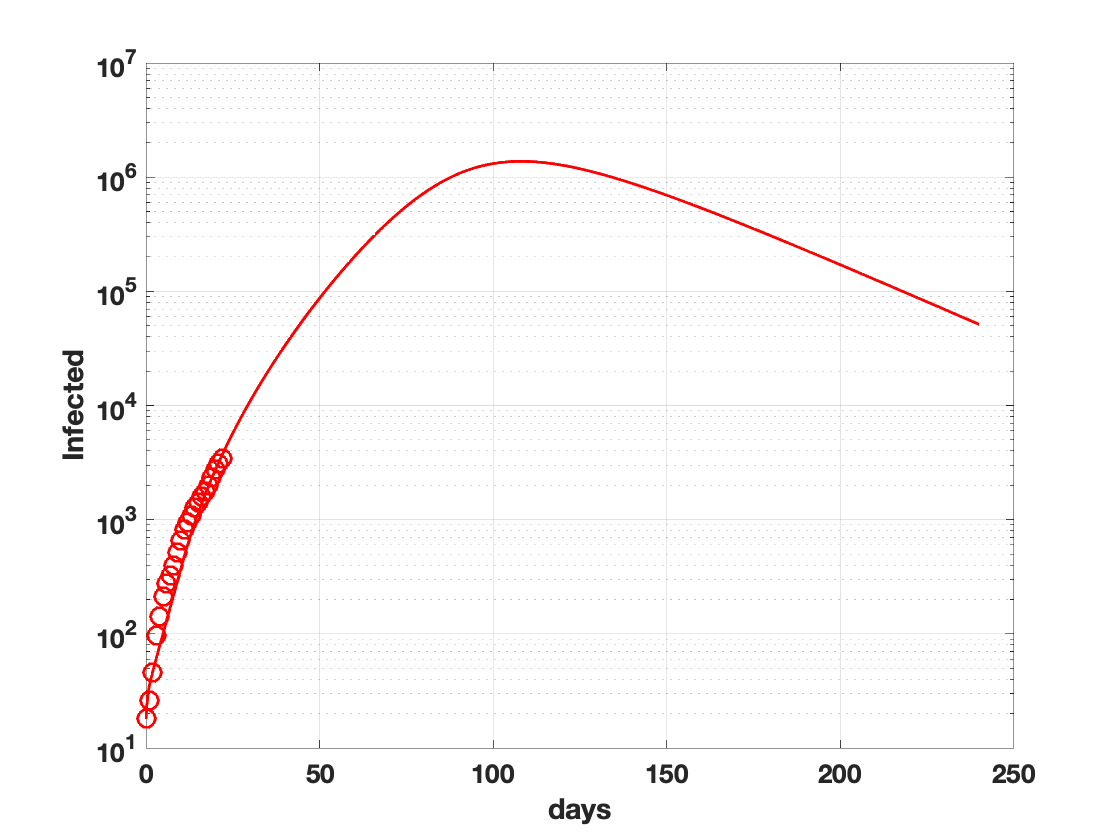}
\includegraphics[width=0.45\textwidth]{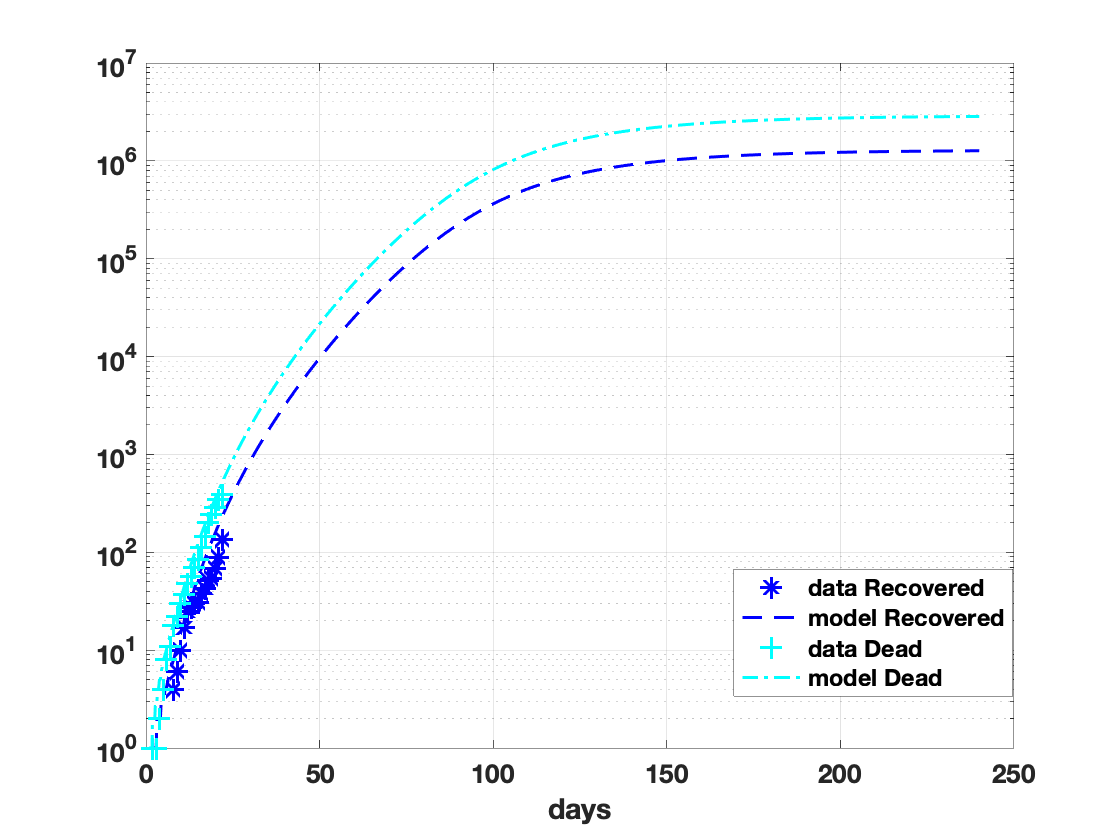}
\caption{Infect, Recovered Dead model based on Emilia Romagna region data (SEIRD(rm)).}
\label{fig:ERMod1}
\end{figure}

\begin{figure}[hbtp]
\centering
\includegraphics[width=0.45\textwidth]{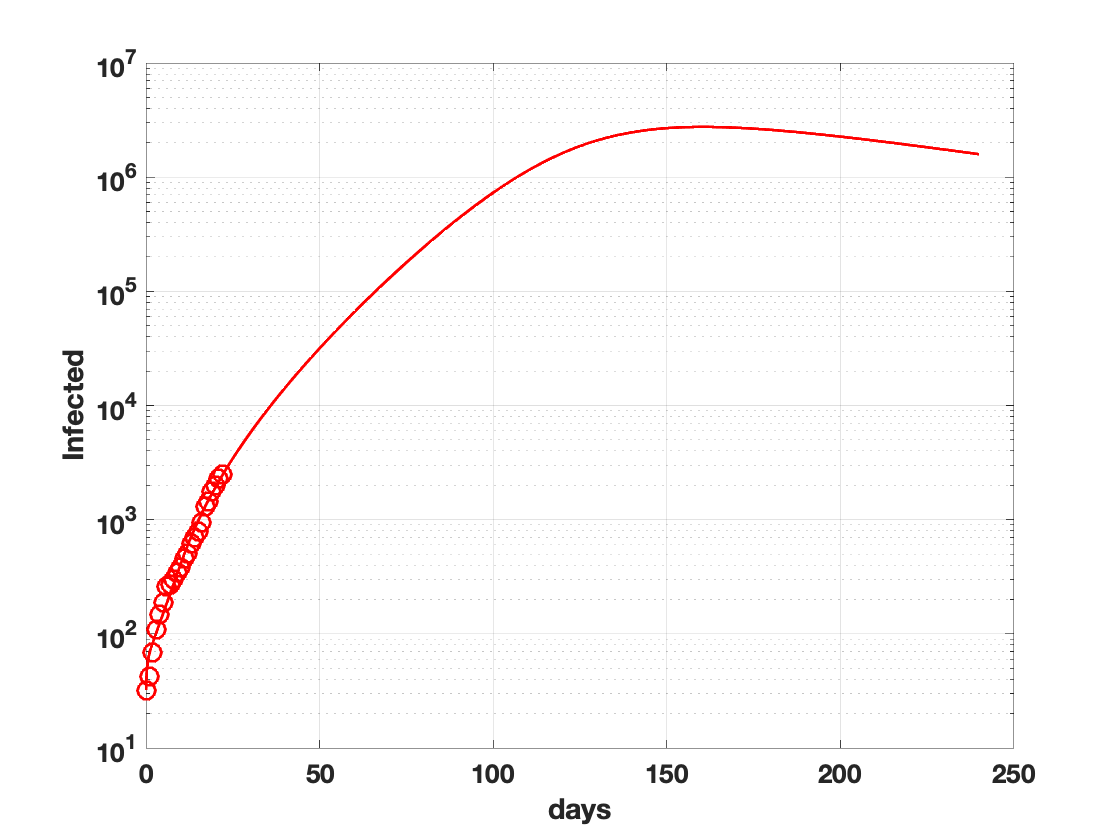}
\includegraphics[width=0.45\textwidth]{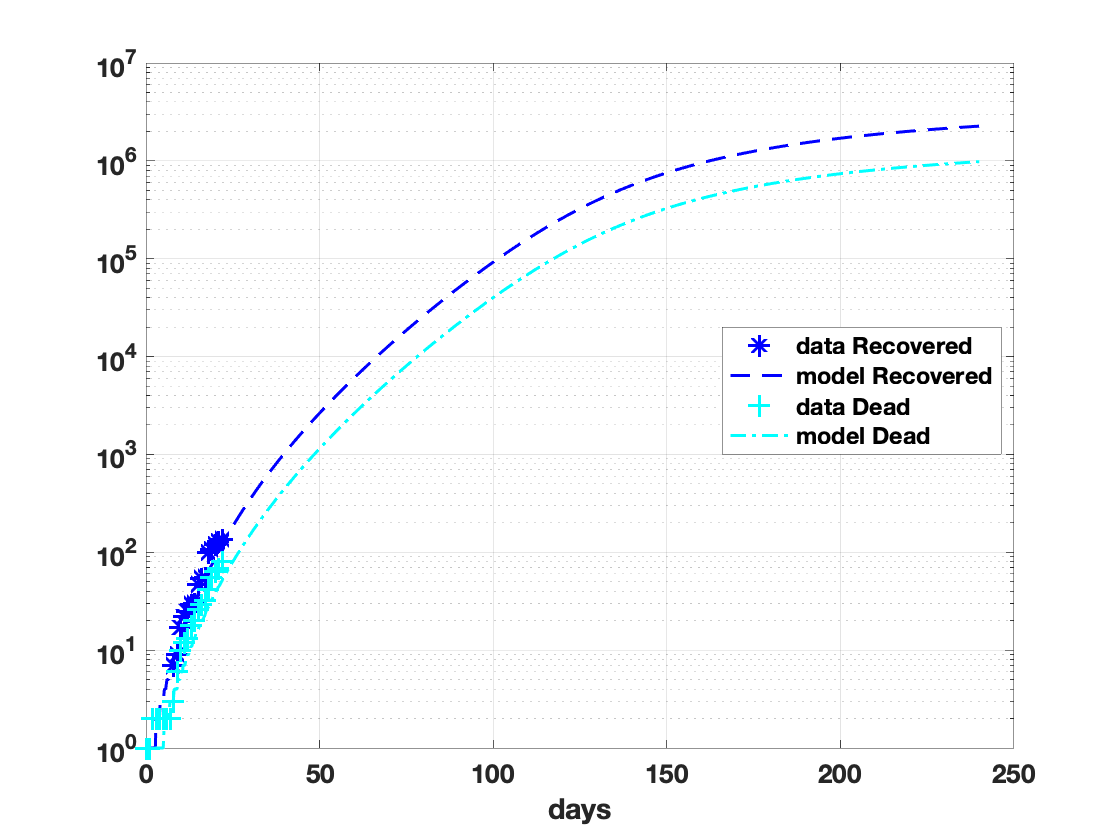}
\caption{Infected, Recovered Dead model based on Veneto region data (SEIRD(rm)).}
\label{fig:VeMod1}
\end{figure}

In order to improve the data fit we split the parameter  identification step into the following  two phase process:
\begin{itemize}
\item {\bf Phase 1} Identification of the parameters of the standard SEIRD model using a the data subset S1, (i.e.  up to time $t_0=10$).
\item  {\bf Phase 2} Identification of the parameters of  SEIRD(rm) model using all the measures available and modelling the infection rate as in \eqref{eq:beta_t}.  The value of the infection rate is constant when $t \leq t_0$ otherwise it has decreasing values. 
\end{itemize}
The parameters obtained from the of the Phase 2 of each region are reported in table \ref{tab:ParsRM}.
%-----------------------------------------------------------------------------     
\begin{table}[!htpb]
\begin{center}
\begin{tabular}{c|ccccc}
Region & $\alpha $ & $\gamma_R $ & $\gamma_D$  \\
\hline
% Lombardia & 0.21 &  0.062 &  0.039  \\
 Lombardia & 0.41 & 0.040 & 0.025 \\
 Veneto & 2.5 & 0.0069 & 0.0030 \\
 Emilia Romagna &  1.35 & 0.011 & 0.025 \\
  \hline
\end{tabular}
\caption{SEIRD(rm) parameters.}
\label{tab:ParsRM}
\end{center}
 \end{table}
%--------------------------------------------------------------
Representing the infection rate as a time dependent  function causes the change of the 
Reproduction parameter (R0) into a time dependent Reproduction function defined as follows:
\begin{equation}
R_t = \frac{\beta(t)}{\gamma_D+\gamma_R}
\label{Rt}
\end{equation}
The plots of $R_t$ for the three regions are reported in figure \ref{fig:Rt}.
In table \ref{tab:Nums1} we report the peak values of the IRD populations
for the  proposed SEIRD(rm) model. 
%------------------------------------------------------------
\begin{table}[!htbp]
\begin{center}
\begin{tabular}{c|cc|cc|cc}
Data &   \multicolumn{2}{c|}{Peak Infected} & \multicolumn{2}{c|}{Peak Recovered}  & \multicolumn{2}{c}{Peak Dead} \\
  Region &            day & Number (\%) &  day &Number (\%) &  day& Number (\%) \\
 \hline
%Lom  &   102 & 1997792 (19.8) & 120 & 3424342 (34.0) & 120 & 2180176 ( 21.6) \\ %17.3
%Lombardia & 106 & 1667171 ( 16.5) & 240 & 5092815 ( 50.5) & 240 & 3145992 ( 31.2) \\ %19_3
Lombardia & 107 & 1615368(16.0\%) & 240 & 5047743(50.1\%) & 240 & 3116082(30.9\%) \\ % \20_3
% Emilia Romagna &   108 & 1376090(30.9) & 240 & 1268228(28.4) & 240 & 2842681(63.7) \\
% Emilia Romagna & 109 & 1440499(32.3) & 240 & 1258678(28.2) & 240 & 2882706(64.6)  \\% 19 Mar
 Emilia Romagna & 109 & 1465451(32.9\%) & 240 & 1260782(28.3\%) & 240 & 2890963(64.8\%) \\ %20 Mar
%Veneto &160 & 2755129(56.2) & 240 & 2263761 (46.1) & 240 & 981688 ( 20.0) \\
%Veneto & 160 & 2823727(57.6) & 240 & 2216958 (45.2) & 240 & 975205 ( 19.9) \\
Veneto &  160 & 2833884(57.8 \%) & 240 & 2208955(45.0\%) & 240 & 979229(20.0\%) \\ %20 Mar
  %                &       69 & 2673967 ( 26.5) & 120 & 6046273 ( 60.0) & 120 3754677 ( 37.2) \\
%Valore residui: I=2.149866e-01 R=1.693460e+00 D=4.808612e-01  
 \hline
 %S2
 \hline 
 \end{tabular}
 \caption{Infected, Recovered Dead values SEIRD(rm) model}
 \label{tab:Nums1}
 \end{center}
 \end{table}
%------------------------------------------------------
%
We observe that the Infected people should reach its peak 
on June 21-22    in Lombardia and Emilia Romagna while Veneto requires the longest time (August 1st 2020) and it seems
to have the largest number of infected people. However,  we highlight that Veneto  
applies a different testing policy with respect to the other regions which tests a larger part of population.
%applies virus tests to almost
%the entire population while the other regions test only symptomatic people. 
Therefore 
the Infected people measured in Veneto contain a greater number of individuals  who do not have symptoms but can spread the infection. 
Following a recent study on Wuhan Covid outbreak \cite{Lieabb3221}  the undocumented infectious can have a great influence in the virus spread. 
These undocumented infectious often experience mild, limited or no symptoms and hence go unrecognized, and, depending on their contagiousness and numbers, can expose a
far greater portion of the population to virus than would otherwise occur.\\
%On the contrary, Infected data of  Emilia Romagna and Lombardia  contain only documented infected individuals with symptoms.
%Therefore we cannot exclude that the pessimistic previsions of Veneto might apply also to the other two regions. \\
The improved modeling properties can be appreciated in population plots reported in figures \ref{fig:LoMod1} , \ref{fig:ERMod1} and \ref{fig:VeMod1}.  
We observe that SEIRD(rm) reproduces the data trends more precisely compared to the SEIRD model.
\\
%Finally we observe a larger rate of dead compared to the recovered people

%--------------------------------------------------------------
%--------------------------------------------------------------
%
%----------------------------------------------------------------------
%\input{NumRes}
\section{Conclusion}
In this paper we proposed a SEIRD model for the analysis of the COVID-19 outbreak diffusion in Italy. In our  new formulation the infection rate coefficient has been adaptively modeled as an inverse function of the time, to take account the restrictions imposed by the Italian Government in the social life on March 8th, 2020. The results obtained by fitting the data of three Italian regions, Lombardia, Emilia-Romagna and Veneto, available since February 24th 2020, show a very good fit to the data and give a  prediction on the behaviour of Infected-Re individuals. 
The forecasts about the maximum infection spread report quite  homogeneous results for both Lombardia and Emilia Romagna (about June 20th), while Veneto has its infection peak around the end of July 2020, probably due to  different testing modalities.\\
We highlight that it is only 12 days  since restrictions started in Italy and, maybe in the next few days,  the effects of such measures will become more evident, hopefully causing a further decrease in the infection trend. In this case, the previsions shown in this paper should be updated by 
introducing a new time $t_1$ at which the the decreasing slope
of  $\beta_t$ should change, for example by estimating the parameter $\rho$ in \eqref{eq:beta_t} with new data.\\
The proposed model is flexible and we believe it could be easily adapted to  monitor various  infected areas with  different restriction policies.
%------------------------------------------

%-----------------------------------------
%\bibliographystyle{natbib}
%\bibliography{biblio_covid}
\end{document}